\documentclass[hyper]{JHEP} 

\usepackage{epsfig}

\newcommand\fverb{\setbox\pippobox=\hbox\bgroup\verb}

\newcommand\fverbdo{\egroup\medskip\noindent%

            \fbox{\unhbox\pippobox}\ }

\newcommand\fverbit{\egroup\item[\fbox{\unhbox\pippobox}]}

\newbox\pippobox

\title{ D2 to
M2 Procedure for D2-Brane DBI
 Effective Action}
\author{by J. Kluso\v{n}\\
     Department of Theoretical Physics and Astrophysics\\
                   Faculty of Science, Masaryk University\\
Kotl\'{a}\v{r}sk\'{a} 2, 611 37, Brno\\
Czech Republic\\
    E-mail: \email{klu@physics.muni.cz}}

\preprint{\hepth{0807.4054}}

 \abstract{We apply
 the procedure
 that was suggested  in
[arXiv:0806.1639] to the case of
abelian D2-brane  Dirac-Born-Infeld
effective action  and discuss its
limitation. Then  we suggest
 an alternative form of this
  procedure that is based on an
   existence of
interpolating action proposed in
[hep-th/9707113,hep-th/9707011].}

\newcommand{\tB}{\tilde{B}}
\newcommand{\tD}{\tilde{D}}

\newcommand{\calg}{\mathcal{G}}
\newcommand{\calgi}{\left(\calg^{-1}\right)}

\newcommand{\hC}{\hat{C}}
\newcommand{\hg}{\hat{g}}
\newcommand{\hX}{\hat{X}}

\newcommand{\mF}{\mathcal{F}}

\newcommand{\tX}{\tilde{X}}

\newcommand{\hk}{\hat{k}}

\def \bAi{\left(\mathbf{A}^{-1}\right)}

\def \bA{\mathbf{A}}

\newcommand{\mL}{\mathcal{L}}

\begin{document}
\section{Introduction}\label{first}
It was proposed by Bagger and Lambert
in collection of very nice papers
\cite{Bagger:2006sk,Bagger:2007jr,Bagger:2007vi}
and independently by Gustavsson in
\cite{Gustavsson:2008dy} \footnote{For
related works, see
\cite{Mukhi:2008ux,Bandres:2008vf,Berman:2008be,VanRaamsdonk:2008ft,
Morozov:2008cb,Lambert:2008et,Distler:2008mk,Gran:2008vi,Ho:2008bn,
Gomis:2008cv,Bergshoeff:2008cz,Hosomichi:2008qk,Papadopoulos:2008sk,
Gauntlett:2008uf,Papadopoulos:2008gh,Ho:2008nn,Gomis:2008uv,Benvenuti:2008bt,
Ho:2008ei,Morozov:2008rc,Honma:2008un,Fuji:2008yj,Ho:2008ve,
Song:2008bi,Jeon:2008bx,Li:2008ez,Hosomichi:2008jd,Banerjee:2008pda,Lin:2008qp,
DeMedeiros:2008zm,Gustavsson:2008bf,FigueroaO'Farrill:2008fz,Bandres:2008kj,
Park:2008qe,Passerini:2008qt,Gomis:2008be,Ezhuthachan:2008ch,Cecotti:2008qs,
Bergshoeff:2008ix,deMedeiros:2008bf,Blau:2008bm,Blau:2008bp,Furuuchi:2008ki,
Agarwal:2008rr,Bandos:2008fr,Bedford:2008hn,Bagger:2008se,Terashima:2008sy,
Terashima:2008ba,Cherkis:2008qr,Chu:2008qv,Bandres:2008ry,Gomis:2008vc,
Schnabl:2008wj,Zhou:2008sa,Li:2008ya,Kim:2008gn,Pang:2008hw,
AliAkbari:2008rm,Hanaki:2008cu,Verlinde:2008di,Imamura:2008dt,Honma:2008ef,
Bergshoeff:2008bh,Krishnan:2008zm}.
 For study of supergravity duals of
these theories, see
\cite{Aharony:2008ug,Ahn:2008ya,Benna:2008zy,Nishioka:2008gz,
Honma:2008jd,Imamura:2008nn,Minahan:2008hf,Gaiotto:2008cg,Ahn:2008gda,
Arutyunov:2008if,Stefanski:2008ik,Grignani:2008is,Hosomichi:2008jb,
Fre:2008qc,Grignani:2008te,Gromov:2008bz,Ahn:2008gd,Gromov:2008qe,Chen:2008qq,
Garousi:2008ik,Hashimoto:2008iv,Astolfi:2008ji,Ahn:2008aa,Lee:2008ui}.}
following earlier works
\cite{Basu:2004ed,Schwarz:2004yj} that
a certain class of $N=8$
super-conformal theories in three
dimensions are potential candidates for
the world-volume description of
multiple M2-branes in M-theory. These
constructions are based on introducing
of an algebraic structure known as Lie
3-algebra that is needed for closure of
supersymmetry algebra. The metric
versions of the above theories fall
into two classes that depend on whether
the invariant bilinear form in
$3$-algebra space is positive definite
or indefinite. The original theories
proposed by Bagger-Lambert are
Euclidean theories with positive
definite bilinear form while more
recent proposals
\cite{Benvenuti:2008bt,Ho:2008ei}
contain bilinear form that is
indefinite  and these Lie 3-algebras
are known as Lorentzian 3-algebras.

It was claimed that the Lorentzian
3-algebra theories capture the
low-energy world-volume dynamics of
multiple parallel M2-branes. This model
has the required classical symmetries,
but has several unresolved problems. In
particular, the classical theory has
ghosts, $X_{\pm}$. Moreover, the
ghost-free formulation seems directly
equivalent to the non-conformal
D2-brane theory as was argued in
\cite{Bandres:2008kj,Gomis:2008be,Ezhuthachan:2008ch}.
Explicitly, it was shown very clearly
in \cite{Ezhuthachan:2008ch} how it is
possible-starting from $N=8$ SYM-
systematically and uniquely recovery
the theory
\cite{Bandres:2008kj,Gomis:2008be}.

Since the analysis presented in
\cite{Ezhuthachan:2008ch}
 is very nice and interesting it certainly deserves
further study. In fact, since $2+1$ dim
$N=8$ SYM theory describes low-energy
dynamics of $N$ D2-branes one can ask
the question whether it is possible to
extend this analysis
\cite{Ezhuthachan:2008ch}
 \footnote{In
what follows we call this analysis as
EMP procedure.}
 when we take
non-linear corrections into account.
 As the first  step in this
direction we try to apply  EMP
procedure
to the  case of single
Dirac-Born-Infeld (DBI) action for
D2-brane.

We start our analysis with the
remarkable form of $2+1$ dimensional
action that was proposed long ago in
\cite{Lozano:1997ee,Ortin:1997jh}. This
action is an interpolating action
 that-after appropriate integration of some
world-volume fields-either describes
D2-brane DBI effective action in
massive Type IIA supergravity or the
directly dimensional reduced gauged
M2-brane action.  We
 show that in linearized level this
action is equivalent to the abelian
form of the action given in
\cite{Ezhuthachan:2008ch} and hence can
be considered as the starting point for
non-linear generalization of EMP
procedure. On the other hand we argue
that naive application of EMP procedure
in this action leads to  a puzzle.
Explicitly, we argue that there is a
unique ground state of this new action
with infinite coupling constant. This
is different from we would expect since
M2 to D2-reduction is based on a
presumption that the vacuum expectation
value of $\left<X_+\right>$ can take
arbitrary constant value. Equivalently,
we would expect an infinite  number of
ground states that differ by vacuum
expectation values of $X_+$.

In order to resolve this problem we
suggest that the natural object for the
definition of non-linear EMP procedure
is gauged M2-brane action.
 More precisely, it is
well known that  in the case of IIA
supergravity it is possible to
introduce non-zero cosmological
constant proportional to $m^2$ with $m$
a mass parameter \cite{Romans:1985tz}.
Such backgrounds are essential for the
existence of D8-branes whose charge is
proportional to $m$
\cite{Bergshoeff:1996ui}. The action
for massive 11-dimensional theory has
the same contain as the massless one
\footnote{We use  the following notation for the hats.
Hats on target space fields
indicate that they are 11-dimensional.}
\begin{equation}
\hg_{MN} \ , \quad \hC_{MNK} \ , \quad
M,N,K=0,\dots, 11 \ .
\end{equation}
The action for these fields
 is manifestly  11-dimensional
Lorentz covariant but it does not
correspond to a proper 11-dimensional
theory because, in order to write down
the action, we need to introduce an
auxiliary non-dynamic vector field
$\hk^M$ such that the Lie derivatives
of the metric and 3-form potential with
respect to it are zero:
\begin{equation}
\mL_{\hk}\hg_{MN}=0 \ , \quad
\mL_{\hk}\hC_{MNL}=0 \ .
\end{equation}
An existence of this Killing vector is
crucial for definition of massive
M2-brane. In fact, the
 world-volume theory of
massive branes \footnote{These
branes-that propagate in the background
with non-zero cosmological constant-are
called as "massive branes" as opposed
to branes that propagate in a
background with zero mass parameter. It
is clear that all these  branes are
massive in the sense that their
physical mass is nonzero.} was
extensively studied in the past, for
example
\cite{Eyras:1998hn,Bergshoeff:1998bs,Bergshoeff:1997ak,Bergshoeff:1997cf}.
These actions  have as a common
characteristic that they are gauged
sigma models. The gauged isometry is
the same as an isometry that is needed
in order to define the massive
11-dimensional supergravity theory. For
example, the original un-gauged
M2-brane action is the same object as
in the  massless theory, i.e. the
corresponding massless  M2-brane.

Let us again return to generalized EMP
procedure. We argue that it can be
naturally applied for gauged M2-brane
action.  As opposite to the original
EMP procedure, where  the Yang-Mills
coupling constant  vector $g_{YM}^I$ is
replaced with a dynamical field $X_+^I$
we replace the constant Killing vector
 $\hk^M$ with dynamical field
$\hX_+^M$. Then we can easily find
manifestly covariant form of the
generalized action with infinite number
of ground states that differ by vacuum
expectation values of $\hX^M_+$.

It is remarkable that the gauged
isometry that  appears in massive
M2-brane action is related to the gauge
symmetry introduced in
\cite{Ezhuthachan:2008ch}. We hope that
this observation will allow  to find
new geometrical interpretations of
gauge symmetries that were introduced
in
\cite{Bandres:2008kj,Gomis:2008be,Ezhuthachan:2008ch}.

The organization of this paper is as
follows. In the next section
(\ref{second}) we introduce the
interpolating D2-brane action and we
argue that after appropriate
redefinition of world-volume fields it
agrees with the abelian version of
D2-brane action introduced in
\cite{Ezhuthachan:2008ch}.  In section
(\ref{third}) we apply EMP prescription
for gauged M2-brane action and we find
covariant and non-linear version of
M2-brane action that has all desired
properties. In conclusion
(\ref{fourth}) we outline our results
and suggest possible extension of our
work. Finally, in Appendix
\ref{Appendix} we explicitly show that
the dimensional reduction in gauged
M2-brane action leads to the
interpolating action introduced in
section (\ref{second}).

\section{D2-Brane Action}\label{second}
We start with the action that was proposed
in \cite{Lozano:1997ee,Ortin:1997jh}
\begin{eqnarray}\label{Ortinact}
S[X^m,X,&V_\mu &,B_\mu]=
\nonumber \\
&=&-\tau_{M2}
\int d^3\xi  e^{-\Phi}\sqrt{-\det
[g_{\mu\nu}+e^{2\Phi}
F_\mu F_\nu]}+\nonumber \\
&+&\frac{\tau_{M2}}{3!} \int d^3\xi
\epsilon^{\mu\nu\rho}
[C^{(3)}_{\mu\nu\rho}+6\pi\alpha'D_\mu
X\mF_{\nu\rho} +6m
(\pi\alpha')^2V_\mu\partial_{\nu}V_{\rho}]
\ ,
\nonumber \\
\end{eqnarray}
where
\begin{eqnarray}
F_\mu &=& D_\mu X+C^{(1)}_\mu \ ,
\nonumber \\
D_\mu X&=&\partial_\mu X+B_\mu \ ,
\nonumber \\
\mF_{\mu\nu}&=&\partial_\mu
V_\nu-\partial_\nu V_\mu-
\frac{1}{2\pi\alpha'}b_{\mu\nu}
\nonumber \\
\end{eqnarray}
and where
\begin{eqnarray}
g_{\mu\nu}&=&\partial_\mu
X^m\partial_\nu X^n g_{mn} \ , \quad
b_{\mu\nu}=b_{mn}\partial_\mu
X^m\partial_\nu X^n \ , \nonumber \\
C_{\mu\nu\rho}^{(3)}&=&C_{mnk}
\partial_\mu X^m\partial_\nu X^n\partial_\rho X^k \ , \quad
C^{(1)}_\mu=C_m^{(1)}\partial_\mu X^m \
, \nonumber \\
\end{eqnarray}
where $g_{mn},b_{mn}$ are space-time
metric and NS two form field
respectively and where
$C^{(3)}_{mnk},C^{(1)}_m$ are
Ramond-Ramond three and one forms
respectively. Further,
 $X^m \ , m,n=0,\dots,9$ are
world-volume modes that describe embedding
of D2-brane in the  target space-time.
Finally, $\tau_{M2}$ is D2-brane
tension defined as
$\tau_{M2}=\frac{1}{l_s^3}$.

 The action (\ref{Ortinact}) contains
extra fields as opposite to usual DBI
action for D2-brane. Firstly, if we
integrate out $B_\mu$ we obtain
\begin{eqnarray}\label{eqB}
-\frac{e^{\Phi}\sqrt{-g}g^{\mu\nu}F_\nu}{
\sqrt{1+e^{2\Phi}g^{\mu\nu}F_\mu
F_\nu}}+
\pi\alpha'\epsilon^{\mu\nu\rho}
\mF_{\nu\rho}=0 \ ,
\end{eqnarray}
where  we used the
fact that
\begin{eqnarray}
\sqrt{-\det[g_{\mu\nu}+e^{2\Phi}F_\mu
F_\nu]}
=\sqrt{-\det g}\sqrt{1+e^{2\Phi}
g^{\mu\nu}F_\mu F_\nu}\ .
\end{eqnarray}
Then if we insert (\ref{eqB}) into
(\ref{Ortinact}) we obtain an action in the
form
\begin{eqnarray}\label{OrtinactDBI}
S= -\tau_{M2} \int d^3\xi e^{-\Phi}
\sqrt{-\det
[g_{\mu\nu}+2\pi\alpha'\mF_{\mu\nu}]}
+\nonumber \\
+\frac{\tau_{M2}}{3!} \int d^3\xi
\epsilon^{\mu\nu\rho}
(C^{(3)}_{\mu\nu\rho}-6\pi\alpha'C_{\mu}^{(1)}\mF_{\nu\rho}
+6m
(\pi\alpha')^2V_\mu\partial_{\nu}V_{\rho})
\
\nonumber \\
\end{eqnarray}
that is standard form of D2-brane in massive
Type IIA background and that reduces  to
the massless Type IIA background when
$m=0$.

In order to see that the action
(\ref{OrtinactDBI}) is related to
abelian reduction of the action given
in \cite{Ezhuthachan:2008ch}  we take
 following background:
\begin{equation}
g_{mn}=\eta_{mn}  \ , \quad
 \Phi=\Phi_0=\mathrm{const} \ , \quad
C^{(1)}_m=C^{(3)}_{mnk}=0 \ .
\end{equation}
Further, let us impose static gauge
\begin{equation}
X^\mu=\xi^\mu \ , \quad \mu=0,1,2
\end{equation}
 so that
\begin{equation}
g_{\mu\nu}=\eta_{\mu\nu}+\delta_{ij}\partial_\mu
X^i\partial_\nu X^j \ , \quad i,j=3,\dots,9
\ .
\end{equation}
Then in the quadratic approximation the
action (\ref{OrtinactDBI}) takes the form
\begin{eqnarray}\label{OrtinactDBIq}
S[X^m,X,V_\mu,B_\mu]&=&-\tau_{M2}\int d^3\xi e^{-\Phi_0}
\sqrt{-\eta}-\tau_{M2}\int d^3 \xi
\sqrt{-\eta}
[\frac{1}{2}e^{-\Phi_0}\eta^{\mu\nu}
\delta_{ij}\partial_\mu X^i
\partial_\nu X^j+\nonumber \\
&+&\frac{e^{\Phi_0}}{2}
\eta^{\mu\nu}F_\mu F_\nu]+
\int d^3\xi \epsilon^{\mu\nu\rho}
(\pi\alpha' \tau_{M2})D_\mu X \mF_{\nu\rho} \ .
\nonumber \\
\end{eqnarray}
As the next step we introduce
the gauge theory  coupling constant
through the standard relations
\begin{eqnarray}
e^{-\Phi_0}l_s^4\tau_{M2}=\frac{1}{g^2_{YM}}
\ , \quad  \left(\tau_{M2}=
\frac{1}{l_s^3} \ , \quad
2\pi\alpha'=l_s\right) \  \quad
\nonumber \\
\end{eqnarray}
so that after rescaling
\begin{equation}\label{rescaling}
\sqrt{\tau_{M2}} e^{\Phi_0/2}X=\tX \ ,
\quad \sqrt{\tau_{M2}} e^{-\Phi_0/2}X^i=
\tX^i \ , \quad
\frac{1}{l_s^{5/2}}B_\mu=\tB_\mu \ , \quad
l_s^{3/2}\mF_{\mu\nu}=\tilde{\mF}_{\mu\nu}
\end{equation}
the action (\ref{OrtinactDBIq})
 takes the form
\begin{eqnarray}
S[\tX^m,\tX,\tilde{V}_\mu,\tB]&=&-\frac{1}{g^2_{YM}l_s^2}\int d^3\xi
\sqrt{-\eta}- \int d^3 \xi \sqrt{-\eta}
[\frac{1}{2}\eta^{\mu\nu}
\delta_{ij}\partial_\mu \tX^i
\partial_\nu \tX^j+\nonumber \\
&+&\frac{1}{2} \eta^{\mu\nu}
(\partial_\mu \tX+g_{YM}\tB_\mu)
(\partial_\nu
\tX+g_{YM}\tB_\nu)]+\nonumber \\
&+& \int
d^3\xi \epsilon^{\mu\nu\rho}
[\frac{1}{2} \tB_\mu \mF_{\nu\rho} +
\frac{1}{2l_s^{3/2}g_{YM}}
\partial_\mu \tX \tilde{\mF}_{\nu\rho}
]
\nonumber \\
\end{eqnarray}
that has the same form as the abelian
form of the action given in
\cite{Ezhuthachan:2008ch}
\footnote{This is true up to total
derivative term since $\int d^3\xi
\epsilon^{\mu\nu\rho}
\partial_\mu \tX \tilde{\mF}_{\nu\rho}=
\int d^3\xi \partial_\mu [\epsilon^{\mu\nu\rho}
\tX \tilde{\mF}_{\nu\rho}]-
\int d^3\xi \tX\partial_\mu(\epsilon^{\mu\nu\rho}
\tilde{\mF}_{\nu\rho})=
\int d^3\xi \partial (\dots)$.}.
Motivated by this result we perform
the rescaling (\ref{rescaling}) in
the action (\ref{OrtinactDBI}) and
we obtain
\begin{eqnarray}\label{OrtinactDBI2}
S[\tX^m,\tX,\tilde{V}_\mu,\tB_\mu]&=&
-\int d^3\xi \sqrt{-\det\bA_{\mu\nu}} +
\int d^3\xi \epsilon^{\mu\nu\rho}
\frac{1}{2} \tB_\mu
\tilde{\mF_{\nu\rho}} \ ,\nonumber \\
\bA_{\mu\nu}&=&
\frac{1}{l_s^{8/3}g_{YM}^{4/3}}\eta_{\mu\nu}+g_{YM}^{2/3}
l_s^{4/3}
\partial_\mu \tX^i\partial_\nu
\tX^j\delta_{ij}+\nonumber \\
&+&l_s^{4/3}g_{YM}^{2/3}(\partial_\mu \tX+
g_{YM}\tB_\mu)(\partial_\nu\tX+g_{YM}\tB_\nu)
 \ ,
\nonumber \\
\end{eqnarray}
where we ignored term that contributes
to the action as total derivative. Now
we are ready to apply EMP procedure for
(\ref{OrtinactDBI2}). We introduce
$8$-dimensional vector $g_{YM}^I$ as
$g_{YM}^I=(\overbrace{0,\dots,0}^{7},g_{YM}),I=1,\dots,8,
$ and "covariant derivative
$\tD$"
\begin{equation}
\tD_\mu \tX^i=\partial_\mu
\tX^i+g_{YM}^i \tB_\mu \ , \quad  \tD_\mu
\tX=\partial_\mu \tX+g_{YM}\tB_\mu \ .
\end{equation}
 Further, we rewrite
 $g_{YM}^2$ in manifest $SO(8)$ covariant manner
 as $g^2_{YM}=g^I_{YM}g^J_{YM}\delta_{IJ}=
|g_{YM}|^2$
and then we replace
vector $g_{YM}^I$ with dynamical field
$X^I_+$ so that the action
(\ref{OrtinactDBI2})
 takes the
form
\begin{eqnarray}\label{OrtinactDBI4}
&S&[\tX^I,\tX_+^I,\tilde{V}_\mu,\tB_\mu,C^\mu_{I}]=
\nonumber \\
&=&-\int d^3\xi\left( \sqrt{-\det
\bA_{\mu\nu}} +\epsilon^{\mu\nu\rho}
\frac{1}{2} \tB_\mu
\tilde{\mF_{\nu\rho}}
+ C^\mu_I \partial_\mu \tX^I_+\right) \
,
\nonumber \\
\bA_{\mu\nu}&=&\frac{1}{l_s^{8/3}
(X^I_+X^J_+\delta_{IJ})^{2/3}}\eta_{\mu\nu}+(
X^I_+X_+^J\delta_{IJ})^{1/3}
l_s^{4/3}\tD_\mu \tX^I\tD_\nu
\tX^J\delta_{IJ} \ ,  \nonumber \\
\end{eqnarray}
where we introduced auxiliary
field $C^\mu_I$ that renders $\tX^I_+$
non-dynamical.

Let us now analyze some  properties of
the action (\ref{OrtinactDBI4}). We are
mainly interested in the study of the
ground state of this theory that has to
solve the equations of motion that
follow from the action
(\ref{OrtinactDBI4}). We presume that
the ground state is characterized by
following configuration of the
world-volume fields
\begin{equation}\label{grst}
\tX^8_+=v=\mathrm{const} \ , \quad
\tD_\mu \tX^I=0 \ , \quad \tB_\mu=0 \ ,
\quad  \tilde{\mF}_{\mu\nu}=0 \ .
\end{equation}
Firstly, the equation of motion for
$C^\mu_I$ takes the form
\begin{equation}
\partial_\mu \tX^I_+=0
\end{equation}
that is clearly obeyed by the ansatz
(\ref{grst}).
On the other hand the equation of
motion for $X^I_+$ takes the form
\begin{eqnarray}\label{eqX+}
-\frac{1}{2}\frac{\delta \bA_{\mu\nu}}
{\delta
X^I_+}\bAi^{\nu\mu}\sqrt{-\det\bA}
-\epsilon^{\mu\nu\rho}
\frac{1}{2l_s^{1/2}(X^I_+X^J_+\delta_{IJ})^{3/2}}
\partial_\mu \tX
\tilde{\mF}_{\nu\rho}=0 \ .
\nonumber \\
\end{eqnarray}
Since  for the ansatz (\ref{grst}) the
matrix $\bA_{\mu\nu}$ is equal to
\begin{equation}
\bA_{\mu\nu}=\frac{1}{l_s^{8/3}v^{4/3}}\eta_{\mu\nu}
\end{equation}
the equation (\ref{eqX+})
implies
\begin{equation}
\frac{1}{v^{3}}=0  \ .
\end{equation}
In other words the ground state
corresponds to the point $v\rightarrow
\infty$ that implies that there is
unique ground state of the theory. As
we argued in introduction this is not
the same what we want since we would
like to have a theory with infinite
number of ground states that differ by
vacuum expectation values of $\tX_+$.
In order to find solution of this
problem we suggest
 an alternative
procedure how to introduce $X_+^I$ as a
new dynamical variable. In the next
section we present such an alternative
procedure that is based on the fact
that the action (\ref{OrtinactDBI}) can
be considered as dimensional reduction
of massive M2-brane.
\section{Gauged Theory for
M2-brane}\label{third}
 Let us again
consider the action (\ref{OrtinactDBI})
and determine the equations of motion
for $V_\mu$
\begin{equation}\label{eqV}
 \pi m\alpha' \epsilon^{\mu\nu\rho}
(\partial_\nu V_\rho-\partial_\rho
V_\mu)+\epsilon^{\mu\nu\rho}
(\partial_\nu B_\rho-\partial_\rho
B_\nu)=0 \ .
\end{equation}
Inserting (\ref{eqV}) back to
(\ref{OrtinactDBI}) we obtain the
action in the form
\begin{eqnarray}\label{OrtinactDBIdr}
S=-\tau_{M2}\int d^3\xi
\sqrt{-\det (e^{-2/3\Phi}
g_{\mu\nu}+e^{4/3\Phi}F_\mu F_\nu)}+
 \nonumber \\
+\frac{\tau_{M2}}{6} \int d^3\xi
\epsilon^{\mu\nu\rho}
(C_{\mu\nu\rho}^{(3)}-3D_\mu X
B^{(1)}_{\nu\rho} +\frac{6}{m}B_\mu
\partial_\nu B_\rho) \ .
\nonumber \\
\end{eqnarray}
As was shown in \cite{Ortin:1997jh}
(and reviewed in appendix
\ref{Appendix}) this action is very
close to the action that one gets by
direct dimensional reduction of the
massive M2-brane that is also known as
gauged M2-brane action. This action can
be defined in the background with
Killing vector isometry $\hk^M(\hX)$.
Then the
  gauged M2-bane action
takes the form \cite{Ortin:1997jh}
\begin{eqnarray}\label{M2gauged}
S&=&-\tau_{M2}\int d^3\xi \sqrt{-\det
D_\mu \hX^M D_\nu \hX^N \hg_{MN}}
+\nonumber  \nonumber \\
&+& \tau_{M2}\int d^3\xi
\epsilon^{\mu\nu\rho} [D_\mu \hX^M
D_\nu \hX^N D_\rho \hX^K
\hC_{MNK}-\frac{6}{m}B_\mu \partial_\nu
B_\rho] \ ,
\nonumber \\
\end{eqnarray}
where the covariant derivative $D_\mu$
is defined as
\begin{equation}
D_\mu \hX^M=\partial_\mu \hX^M +B_\mu
\hk^M(\hX) \ ,
\end{equation}
where $B_\mu$ is world-volume gauge
field related to the Killing gauge
isometry. To clarify meaning of this
gauged form of the action let us
consider following transformation
\begin{equation}\label{deltahx}
\delta_\eta\hX^M(\xi)=
\hX'^M(\xi)-\hX^M(\xi)=
\eta(\xi)\hk^M(\hX) \ ,
\end{equation}
where $\eta(\xi)$ is a parameter of
gauge transformations. This
transformation immediately implies
following transformation rules of
background fields
\begin{eqnarray}\label{deltag}
\delta_\eta \hg_{MN}=\eta \hk^K
\partial_K \hg_{MN} \ , \quad
\delta_\eta \hC_{KMN}=\eta \hk^L\partial_L
\hC_{KMN} \ , \quad
\delta_\eta \hk^K=\eta \hk^L\partial_L \hk^K
\nonumber \\
\end{eqnarray}
and transformation of covariant
derivative
\begin{eqnarray}
\delta_\eta D_\mu \hX^M=
\eta D_\mu
\hX^L\partial_L\hk^M \ ,
\end{eqnarray}
where we postulate following
transformation rule for gauge field
$B_\mu$
\begin{equation}
\delta B_\mu=-\partial_\mu \eta \ .
\end{equation}
Then
\begin{eqnarray}
\delta (D_\mu \hX^M D_\nu \hX^N
\hg_{MN})= \eta  D_\mu \hX^M(
\partial_M \hk^L \hg_{LN}+
\hg_{ML}\partial_N \hk^L+
\partial_L \hg_{MN})D_\nu \hX^N=0
\nonumber \\
\end{eqnarray}
since
\begin{equation}
\mL_{\hk}\hg_{MN}=0 \ .
\end{equation}
In the same way we obtain that
\begin{equation}
\delta_\eta (D_\mu \hX^M D_\nu \hX^N
D_\rho \hX^K \hC_{MNK})=0 \ , \quad
\mL_{\hk}\hC_{MNK}=0
\end{equation}
hence we see that the action is
invariant under  transformations
(\ref{deltahx}) and (\ref{deltag}).

Having clarified the fact that the
D2-brane action (\ref{OrtinactDBI}) is
related to the gauged M2-brane action
we now introduce  modified EMP
procedure to the action
(\ref{M2gauged}). As the first step in
our construction we will presume an
existence of  adapted system of
coordinates where $\hk^M=const$. This
is always possible to achieve in flat
background
$\hg_{MN}=\eta_{MN},\hC_{MNK}=0$.
Further, in analogy with EMP
prescription, we replace constant
$\hk^M$ with  dynamical field $\hX^M_+$
and add to the action term
$\frac{1}{2}C^\mu_M
\partial_\mu \hX^M_+$ to render this field
non-dynamical.

Further  we rewrite the Wess-Zumino
term in (\ref{M2gauged}) as
\begin{equation}\label{Vin}
\pi\alpha'\epsilon^{\mu\nu\rho} B_\mu
\mF_{\nu\rho}+ (\pi\alpha')^2m
\epsilon^{\mu\nu\rho}V_\mu\partial_\nu
V_\rho \ .
\end{equation}
In fact it is easy to see that now
the equation of motion for $V_\mu$
that follow from (\ref{Vin}) implies
\begin{equation}
\frac{1}{m\pi\alpha'}
 (\partial_\mu
B_\nu-\partial_\nu B_\mu)= -
(\partial_\mu V_\nu-\partial_\nu V_\mu)
\end{equation}
and hence when we insert it back to (\ref{Vin})
 we obtain the last term in (\ref{M2gauged}). Note
also that this expression is invariant
under $\eta$ transformations (up to
total derivative) since
\begin{equation}
\delta (\epsilon^{\mu\nu\rho} B_\mu
\mF_{\nu\rho})= \epsilon^{\mu\nu\rho}
\partial_\mu \eta \mF_{\nu\rho}=
-\eta\epsilon^{\mu\nu\rho}
\partial_\mu\partial_\nu A_\rho+
\eta\epsilon^{\mu\nu\rho}
\partial_\mu\partial_\rho A_\nu=0 \ .
\end{equation}
In summary we derive the action in the
form
\begin{eqnarray}
S&=&-\tau_{M2}\int d^3\xi [\sqrt{-\det
\calg_{\mu\nu}}+\frac{1}{2}\sqrt{-\det
\calg_{\mu\nu}}C_\nu^N\eta_{NM}
\calgi^{\nu\mu}\partial_\mu\hX^M_+
+\nonumber  \\
&+& \frac{\tau_{M2}}{3!}\int d^3\xi
\epsilon^{\mu\nu\rho} [6\pi\alpha'
B_\mu \mF_{\nu\rho}+6m(\pi\alpha')^2
V_\mu
\partial_\nu V_\rho] \ ,
\nonumber \\
\end{eqnarray}
where we  added term $\frac{1}{2}
C^M_\mu \calgi^{\mu\nu}\eta_{MN}\partial_\nu
\hX^N_+$ that renders $\hX_+^M$
constant on-shell and
where we also introduced
"generalized metric" $\calg_{\mu\nu}$
\begin{equation}
\calg_{\mu\nu}= (\partial_\mu
\hX^M+B_\mu \hX^M_+)
\eta_{MN}(\partial_\nu \hX^N+ B_\nu
\hX^N_+) \ .
\end{equation}

Following \cite{Ezhuthachan:2008ch}
 we introduce   field $\hX^M_-$ and add to
the action an expression
\[\frac{1}{2}\sqrt{-\det\calg}
\partial_\mu \tX^M_-\eta_{MN}\calgi^{\mu\nu}
\partial_\nu \tX^N_+ \]
in order the action will be invariant
under  additional shift symmetry
\begin{equation}\label{deltash}
\delta C^M_\mu=\partial_\mu\lambda^M
 \ , \quad  \delta \hX^M_{-}=
\lambda^M \ .
\end{equation}
Then the final form of the action
 takes the form
\begin{eqnarray}
S&=&-\tau_{M2}\int d^3\xi \sqrt{-\det
\calg_{\mu\nu}}\left(1+\right.\nonumber \\
&+& \left.\frac{1}{2}(C_{\mu}^M
-\partial_\mu\hX_{-}^M)
\eta_{MN}
 \calgi^{\mu\nu}
\partial_\nu
\hX^M_+\right)
+\nonumber  \\
&+& \frac{\tau_{M2}}{3!}\int d^3\xi
\epsilon^{\mu\nu\rho} [6\pi\alpha'
B_\mu \mF_{\nu\rho}+6m(\pi\alpha')^2
V_\mu
\partial_\nu V_\rho]
\nonumber \\
\end{eqnarray}
Note also that in order to achieve that
 $\hX_+^M$ is constant on-shell and
that the action possesses additional
shift symmetry we can consider more
general form of the action
\begin{eqnarray}
S&=&-\tau_{M2}\int d^3\xi \sqrt{-\det
\calg_{\mu\nu}}
\sqrt{1+(C_\mu^M-\partial_\mu \hX_-^M)
\eta_{MN}\calgi^{\mu\nu}\partial_\nu \hX^N_+}
-\nonumber  \\
&-& \frac{\tau_{M2}}{3!}\int d^3\xi
\epsilon^{\mu\nu\rho} [6\pi\alpha'
B_\mu \mF_{\nu\rho}+6m(\pi\alpha')^2
V_\mu
\partial_\nu V_\rho]
\nonumber \\
\end{eqnarray}
that can be finally written in a
suggestive form as
\begin{eqnarray}\label{M2final}
S&=&-\tau_{M2}\int d^3\xi \sqrt{-\det
\bA_{\mu\nu}}
-\nonumber \\
&-& \frac{\tau_{M2}}{3!}\int d^3\xi
\epsilon^{\mu\nu\rho} [6\pi\alpha'
B_\mu \mF_{\nu\rho}+6m(\pi\alpha')^2
V_\mu
\partial_\nu V_\rho]
\nonumber \\
\bA_{\mu\nu}&=&
\calg_{\mu\nu}+(C_\mu^M-\partial_\mu \hX^M_-)
\eta_{MN}\partial_\nu \hX^N_+ \ .
\nonumber \\
\end{eqnarray}
Let us now study properties of the
action (\ref{M2final}).  Clearly it
 is invariant
under shift symmetry (\ref{deltash}).
Further, the variation of this action with respect to
$C^M_\mu$ implies
\begin{equation}
\eta_{NM}\partial_\nu \hX^M_+
\bAi^{\nu\mu}\sqrt{-\det\bA}=0
\end{equation}
that  implies $\partial_\nu \hX^M_+=0$.
Let us again presume the ground state
of the theory in the form
\begin{equation}\label{anstM}
B_\mu=\hX^M_-=C^M_\mu=V_\mu=0 \ , \quad
\hX^M_+=v^M \ .
\end{equation}
It is easy to see that the equations
of motion for $\hX^M,B_\mu,C^M_\mu$ and
$\hX^M_-$ are obeyed for this ansatz. Finally,
the problematic equation of motion for
$\hX^M_+$ takes the form
\begin{eqnarray}
& & B_\mu \eta_{MN}(\partial_\nu
\hX^N+B_\nu \hX^N_+)\bAi^{\nu\mu}
\sqrt{-\det\bA}+\nonumber \\
&+&\frac{1}{2}
\partial_\mu[\eta_{MN}\partial_\nu
\hX^N_-\bAi^{\nu\mu}\sqrt{-\det\bA}]=0
\nonumber \\
\end{eqnarray}
that is clearly solved by (\ref{anstM})
for any $v^M$. Finally, let us impose
the static gauge in the following form
\begin{eqnarray}
\hX^\mu=\xi^\mu \ , \quad  \mu \ , \nu=0,1,2 \nonumber \\
C_\mu^\nu=0 \ , \quad
\hX^\mu_+=\hX^\mu_-=0
\nonumber \\
\end{eqnarray}
so that
the matrix $\bA_{\mu\nu}$ takes the form
\begin{eqnarray}
\bA_{\mu\nu}&=&
\eta_{\mu\nu}+(\partial_\mu \hX^I+B_\mu
\hX^I_+) (\partial_\nu \hX^J+B_\nu
\hX^J_+)\delta_{IJ}+
\nonumber \\
&+&(C_\mu^I-\partial_\mu
\hX_-^I)\delta_{IJ}
\partial_\nu \hX^J_+ \ , \quad I,J=1,\dots,8 \ . \nonumber \\
\end{eqnarray}
Then the action  up to quadratic approximation
can be written as
\begin{equation}
S=-\tau_{M2}\int d^3\xi
\sqrt{-\eta}-\tau_{M2}
\int d^3\xi \sqrt{-\det\eta}
\mL \ ,
\end{equation}
where the Lagrangian density takes the form
\begin{eqnarray}
\mL&=&\frac{1}{2} \eta^{\mu\nu}
(\partial_\mu \hX^I+B_\mu\hX^I_+)
\delta_{IJ} (\partial_\nu
\hX^J+B_\nu\hX^J_+)\delta_{IJ}
+\nonumber \\
&+&\frac{1}{2}
\eta^{\mu\nu}(C_\mu^I-\partial_\mu \hX^I_-)
\partial_\nu \hX^J_+\delta_{IJ}-
\nonumber \\
&-&\epsilon^{\mu\nu\rho}
(\pi\alpha' B_\mu \mF_{\nu\rho}+
m(\pi\alpha')^2V_\mu\partial_\nu V_\rho) \ .
\nonumber \\
\end{eqnarray}
that is again very close to the abelian
form of the action given in
\cite{Ezhuthachan:2008ch} and provides
further support of our construction.


\section{Conclusion}\label{fourth}
Let us summarize our results. We
studied EMP procedure for
Dirac-Born-Infeld action for D2-brane
and we found its limitation. Then we
suggested an alternative form of this
procedure that  is based on a
formulation of gauged M2-brane action.
This fact however implies that the
theory should be defined in background
with non-zero mass parameter $m$ and
this observation certainly deserves
better understanding and more detailed
 study. Further, it would be also
interesting to develop BRST Hamiltonian
treatment of the action (\ref{M2final})
and compare it with the similar
analysis that was given in
\cite{Benvenuti:2008bt}. Finally, it
will be extremely interesting to see
whether there exists an non-abelian
extension of the action
(\ref{M2final}). We hope to return to
these problems in future.
\\
\\
{\bf Acknowledgement}

This work
 was supported  by the Czech Ministry of
Education under Contract No. MSM
0021622409.

\begin{appendix}
\section{Direct Dimensional Reduction for
Massive M2-brane}\label{Appendix} In
this appendix we show that the gauged
M2-brane action upon direct dimensional
reduction in the direction $X$
associated to the gauged isometry
reduces to the action
(\ref{OrtinactDBIdr}). To begin with we
choose coordinates that are adapted to
the isometry so that
$\hk^M=\delta^{Mx}$ and we split eleven
coordinates $\hX^M$ into the ten
$10$-dimensional $X^m,m=0,\dots,9$ and
the extra scalar $\hX^{x}\equiv X$.
Using the relations between the
$11$-dimensional and $10$-dimensional
fields
\begin{eqnarray}\label{ansA}
\hg_{xx}=e^{\frac{4}{3}\Phi} \ , \quad
\hg_{mx}=e^{\frac{4}{3}\Phi} C^{(1)}_m
\ ,
\nonumber \\
\hg_{mn}=e^{-\frac{2}{3}\Phi}g_{mn}+
e^{\frac{4}{3}\Phi}C^{(1)}_m C^{(1)}_n
\ ,
\nonumber \\
\hC_{mnk}=C^{(3)}_{mnk} \ , \quad
\hC_{mnx}=B_{mn} \  \nonumber \\
\end{eqnarray}
 it is straightforward to see that
\begin{equation}\label{bAA}
\bA_{\mu\nu}=e^{-\frac{2}{3}\Phi}
\partial_\mu X^m\partial_\nu X^n g_{mn}+
e^{\frac{4}{3}\Phi}(\partial_\mu
X+B_\mu+C^{(1)}_\mu) (\partial_\nu
X+B_\nu +C^{(1)}_\nu) \ .
\end{equation}
Then if we insert (\ref{bAA}) together
with (\ref{ansA}) into the action
(\ref{M2gauged}) we easily obtain that
it  reduces to the  action
(\ref{OrtinactDBIdr}).
\end{appendix}

\end{document}